\pdfoutput=1

\documentclass[11pt]{article}

\usepackage[]{acl}

\usepackage{times}
\usepackage{latexsym}
\usepackage{graphicx}

\usepackage[T1]{fontenc}
\usepackage{verbatim}
\usepackage[utf8]{inputenc}

\usepackage{microtype}

\title{Enhancing Collaborative Filtering Recommender with Prompt-Based Sentiment Analysis}

\author{
  Elliot Dang\\
  New York University \\
  \texttt{yd1008@nyu.edu}
  \And
  Zheyuan Hu\\
  New York University \\
  \texttt{zh2095@nyu.edu} 
  \And
  Tong Li\\
  New York University \\
  \texttt{tl2204@nyu.edu} 
}

\begin{document}
\maketitle
\begin{abstract}
Collaborative Filtering(CF) recommender is a crucial application in the online market and e-commerce. However, CF recommender has been proven to suffer from persistent problems related to sparsity of the user rating that will further lead to a cold-start issue. Existing methods address the data sparsity issue by applying token-level sentiment analysis that translate text review into sentiment scores as a complement of the user rating. In this paper, we attempt to optimize the sentiment analysis with advanced NLP models including BERT and RoBERTa, and experiment on whether the CF recommender has been further enhanced. We build the recommenders on the Amazon US Reviews dataset, and tune the pretrained BERT and RoBERTa with the traditional fine-tuned paradigm as well as the new prompt-based learning paradigm. Experimental result shows that the recommender enhanced with the sentiment ratings predicted by the fine-tuned RoBERTa has the best performance, and achieved 30.7\% overall gain by comparing MAP, NDCG and precision at K to the baseline recommender. Prompt-based learning paradigm, although superior to traditional fine-tune paradigm in pure sentiment analysis, fail to further improve the CF recommender.

\end{abstract}

\section{Introduction}
Collaborative filtering (CF) is a techniques widely used by recommender systems. As one of the two categories of CF, explicit CF exploit feedback such as numerical ratings of an existing user’s community to predict which items the current user probably like most \cite{Schafer2007}. Explicit CF perform well as long as there is sufficient rating information. However, their effectiveness deteriorates if there exist insufficient ratings, which is known as data sparsity, and data sparsity would further lead to a cold start issue \cite{BOBADILLA2013109}. Meanwhile, the rating criteria differ for each user, and inconsistent rating criteria would weaken the reliability of the recommendation. 

One possible way to reduce the data sparsity and inconsistent rating criteria is to integrate the text reviews into the recommender system, and apply sentiment analysis on the reviews \cite{GARCIACUMBRERAS20136758}. The numerical scores decoded from the sentiment analysis can be used as a supplement to the ratings with a uniform standard, and then filled into an expanded user-item matrix to enhance the recommender. 

Recently, prompt learning (pretrain - prompt -predict) as a new paradigm in Natural Language Processing(NLP), has shown its potential to outperform the fine-tune - pretrain paradigm, and has been proved successful in a wide range of NLP tasks, including sentiment analysis. \cite{DBLP:journals/corr/abs-2107-13586}. 

In this paper, we investigated whether the state-of-the-art techniques in NLP would help with the sentiment analysis to further diminish the impact of data sparse issue. We developed three different recommenders, the baseline recommender fit on the original sparse dataset, the one enhanced by sentiment analysis with finetune-pretrain paradigm, and the one enhanced by sentiment analysis with prompt learning paradigm. We compared the above recommenders based on the evaluation metrics that measured both predictive accuracy as well as the user experience, including MAP at K, NDCG, and Precision at K.

\section{Related Work}

\textbf{Early Solution} One of the early solutions to overcome the cold-start recommendation issue is to exploit social tags of the users \cite{10.3844/jcssp.2014.1166.1173}. Tagging indeed has the potential of enhancing finding similar users, but it is also based on the naive assumption that users are interested with the items that were annotated.
\newline
\textbf{Transition to sentiment analysis} According to \cite{GARCIACUMBRERAS20136758}, a sentiment analysis approach can be applied to textual reviews in order to infer users’ preferences and such preferences can be subsequently mapped into some numerical ratings that the CF algorithm relies on. Ricci \cite{Ricci2015} also discussed the cold start challenge in recommender systems, and proposed that applying sentiment analysis to be one of the solutions.
\newline
\textbf{Token level sentiment analysis} In this case, Osman and Nurul \cite{10.1371/journal.pone.0248695} has developed a CF recommender system that integrated with sentiment analysis on text reviews as an enhancement. They designed an algorithm that translates text reviews into sentiment scores, and then supplement the missing ratings with the sentiment scores. However, the algorithm is simply based on the occurrence of positive and negative terms, so it only captures the sentiment at the token level, but fail to retrieve the contextual sentiment information at the sentence level. 
\newline
\textbf{Pretrained language model} With the arrival of pre-trained language models(PLMs) such as Bidirectional Encoder Representations from Transformers (BERT) \cite{devlin2018bert}, Sentiment analysis from text data has undergone a colossal transformation. Shivaji and Manit \cite{Alaparthi2021} concluded that BERT has the undisputed superiority in sentiment classification from text, by running experiments on the IMDB movie review dataset and comparing the performance of the unsupervised SentiWordNet, Logistic Regression, LSTM, and Bert-Base pretrained on Wikipedia and Bookcorpus. 
\newline
\textbf{Prompt Learning}
Recently, prompt-based learning has been proved successful in multiple NLP tasks, including sentiment analysis \cite{DBLP:journals/corr/abs-2107-13586}. Prompting shortens the gap between the objectives for the pre-training stage and the objectives for the downstream tasks by designing a template that transforms the downstream inputs to a certain format, so that enables the input to fit the PLM properly.

\section{Methodology}

\subsection{Data}
The original dataset is the Amazon US Reviews dataset with the category "Video", which contains both ratings and textual reviews, and can be accessed through the HuggingFace API. The preprocessed dataset contains roughly 280K items, each represented by the following 4 columns required for running sentiment analysis and building CF recommender: 
\begin{itemize}
\item customer\_id: Random identifier represents customer who purchased the video.
\item product\_id: The unique Product ID the review pertains to.
\item star\_rating: The 1-5 numerical rating of the video.
\item review\_headline: The content of the textual review.
\end{itemize}

In order to obtain a sparse dataset to simulate the data sparsity issue, we randomly dropped 40\% of the user rating, so the dataset contains only 60\% of the true user rating. Further operations (e.g.create user-item matrix, obtain sentiment ratings) and experimentation are achieved with this preprocessed sparse dataset.

\subsection{Explicit Collaborative Filtering}

Collaborative Filtering (CF) model is widely used for RecSys, which usually takes a model-based approach. This approach strives to construct latent representations for User and Item and uses Alternating Least Squares (ALS) to learn the representations. ALS is based on the concept of Matrix Factorization and iteratively solves for the best lower-dimensional latent representations. Each rating, or explicit user-item interaction, represents the preference of the user to the item, where a higher rating usually suggests a higher preference. CF recommends items to users by constructing a expected score for the missing user-item interactions by using the latent representations, and the scores reflect the magnitude of confidence for the recommendation. 

\subsection{Fine-Tuned PLM}
Masked language model (MLM) has been proved powerful for sentiment analysis and its generalizability for the downstream tasks has been guaranteed by previous studies as well. As a first attempt that applies both PLM and prompt learning to enhance the CF recommender, we consider BERT\cite{devlin2018bert} and RoBERTa \cite{DBLP:journals/corr/abs-1907-11692} to be the best choices of the PLM, since they are both representative MLMs, and are feasible enough for the prompt-based learning paradigm. Specifically, we decide to use the base version for both BERT and RoBERTa, that are pretrained on the Wikipedia and Bookcorpus dataset, and contain 110M and 125M parameters respectively, due to the performance-runtime trade-off. 

Since sentiment analysis is basically a sequence classification task, where the numeraical ratings from 1-5 can be considered as 5 labels, we add an multilayer perceptron(MLP) as the classification head of the PLMs that maps the last hidden layer to the label space with the softmax activation function. We then fine tune the PLMs on the 60\% of the dataset where the rating exist, with the embedded textual reviews as the input and corresponding numerical ratings as the target label. Once we obtain the fine-tuned PLMs, we complement the 40\% whose rating has been dropped with the sentiment scores predicted by the fine-tuned PLMs. Eventually, we can build CF recommenders that are enhanced by sentiment analysis with the fine-tuned PLMs with 60\% of the true user rating and 40\% of the sentiment rating.

\subsection{Prompt Learning}
\begin{figure}[ht]
\hspace{0cm}  
\includegraphics[scale=0.1]{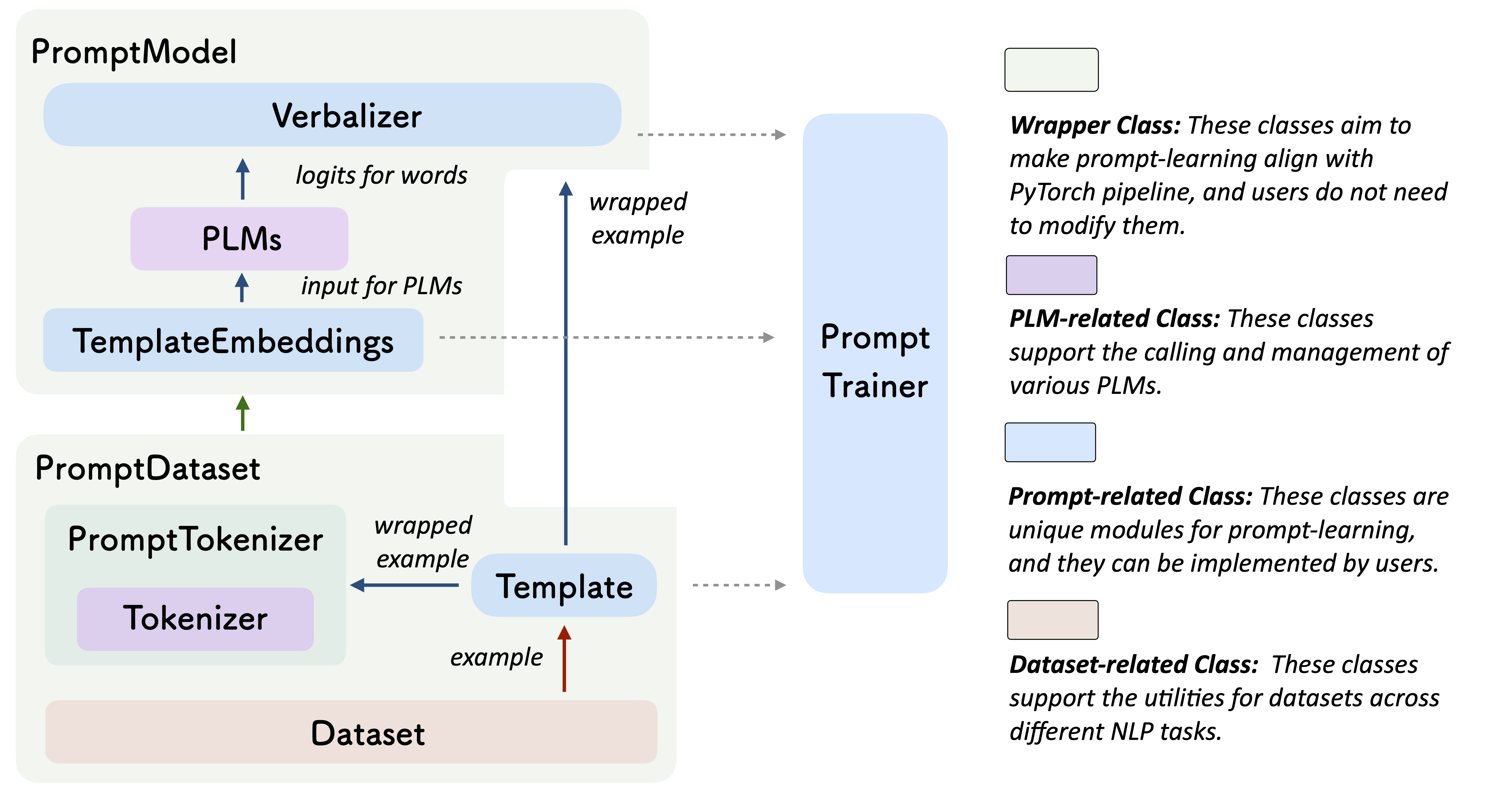}
\caption{The overall workflow of prompt learning with OpenPrompt}
\label{fig1}
\end{figure}

We implement manual prompt design by basically following the pipeline proposed by OpenPrompt \cite{DBLP:journals/corr/abs-2111-01998}, which is a standard, and flexible framework. Figure \ref{fig1} illustrates the overall workflow of OpenPrompt. In order to construct the PromptModel and PromptDataset, we first define the following object:
\begin{itemize}
\item PLMs - The backbone model for the prompt-based learning. Here we choose BERT base and RoBERTa base with a sequence classification head on top. 
\item Verbalizer: A verbalizer class maps original labels to label words in the vocabulary. In this case, we have 5 labels representing the numerical ratings,so the verbalizer will map \{1,2,3,4,5\}to the word vocabulary \{awful, bad, fair, good, wonderful\}, respectively.
\item Template - A modifier that converts the input text into certain format. Here the format is designed to be "Overall, it was a [x] movie", and the verbalizer is defined as above. For example, if we have a user review "I love this movie" with a 4 star rating, the Template would then replace "[x]" with the word "good" mapped from the numerical rating 4, that is, the review will be converted to "Overall, it was a good movie" by the Template.
\end{itemize}

Once the PLM, Template, and Verbalizer has been well defined, the text reviews in the original dataset will be modified by the template and then be tokenized with the tokenizer pretrained by the PLM to obtain the PromptDataset. Meanwhile, the PromptModel object that practically participates in training and inference will be constructed by combining the Verbalizer, the PLM, and the Template together. The Trainer module will be responsible for tuning the PLM and prompts simultaneously with the same training configuration as the pretrain-finetune paradigm.

Similarly, we train on 60\% of the dataset where the rating exists and predict the prompt-based sentiment rating for the remaining 40\% where rating has been dropped.

\section{Experiments}
\subsection{Experiments Setup}

We plan to compare the RecSys performance trained with five different datasets: \textbf{SPARSE} the baseline, \textbf{SENT-BERT} that uses BERT, \textbf{SENT-ROBERTA} that uses RoBERTa, \textbf{SENT-PROMPT-BERT} that incorporates BERT with Prompt Learning, and finally \textbf{SENT-PROMPT-ROBERTA} that combines RoBERTa with Prompt Learning. The schemes used to create these datasets are explained in Section 3. We use the same validation set the measure the performance.

\subsection{Evaluation Results}
\begin{table}[ht]
\centering
\begin{tabular}{||c c c||} 
 \hline 
 Model & Accuracy & F1 \\ [0.5ex] 
 \hline\hline 
 BERT & 0.7418 & 0.7268 \\ [0.5ex]
 \hline
 RoBERTa & 0.7662 & 0.7551  \\ [0.5ex]
 \hline 
 BERT-Prompt & 0.7600 & 0.7641  \\ [0.5ex]
 \hline 
 RoBERTa-Prompt & \textbf{0.7832} & \textbf{0.7658}  \\ [0.5ex]
 \hline
\end{tabular}
\caption{Model performance on 5-label sentiment classification}
\label{tab1}
\end{table}

The four PLMs have been tuned to achieve the accuracy and F1 score for sentiment classification as shown in Table \ref{tab1}. RoBERTa outperforms BERT in both accuracy and F1 as expected. Meanwhile, prompt-learning paradigm has demonstrated its superiority over the traditional finetune-pretrain paradigm for both BERT and RoBERTa. We are then interested in whether the CF recommender further benifits from prompt-learning.

\begin{table*}[ht]
\centering
\begin{tabular}{||c c c c c||} 
 \hline 
RecSys & MAP & NDCG@30 & P@30 & Avg. Imp\% \\ [0.5ex] 
 \hline\hline 
 SPARSE & 0.4598 & 0.4906 & 0.0304 & ---\\ [0.5ex]
 \hline
 SENT-BERT & 0.4854 & 0.5409 & \textbf{0.0518} & 28.74\% \\ [0.5ex]
 \hline 
 SENT-ROBERTA & \textbf{0.5999} & \textbf{0.6403} & 0.0399 & \textbf{30.74}\% \\ [0.5ex]
 \hline 
 SENT-PROMPT-BERT & 0.5832 & 0.6258 & 0.0392 & 27.78\% \\ [0.5ex]
 \hline
 SENT-PROMPT-ROBERTA & 0.5770  & 0.6196 & 0.0389 & 26.58\% \\ [0.5ex]
 \hline
\end{tabular}
\caption{Evaluation results for recommenders}
\label{tab:res}
\end{table*}

We adopt Spark.MLlib to train an Alternating Least Squares (ALS) for our Collaborative Filtering RecSys with explicit feedback (rating). We evaluate our recommendation performance on three different metrics: Mean Average Precision (MAP), Normalized Discounted Cumulative Gain at K (NDCG@K), and Average Precision at K (P@K). These are commonly used metrics for recommendation systems, which compute metrics between the ranked list of recommendation, and the set of group truth items (items that the user actually consume), where ranking is based on the score produced by the RecSys. 

Suppose we have $M$ users, and $U = \{u_1, u_2, ..., u_M\}$ be the set of all users. For each user $u_i$, it has a set of $N_i$ ground truth items $D_i = \{d_1, d_2, ..., d_{N_i}\}$. And also a ranked list containing $Q_i$ recommended items $R_i = [r_1, r_2, ..., r_{Q_i} ]$. Define a relevance function $rel_D(r) = 1 \ \ if \ \ r \in D, 0$ otherwise, and let $n = \min(\max(Q_i, N_i), K)$, $IDCG(D, K) = \sum_{j=1}^{min(|D|, K)}\frac{1}{\log(j+1)}$.The metrics are defined as follows:
    \begin{itemize}
        \item $P@K = \frac{1}{M} \sum_{i=1}^M \frac{1}{K} \sum_{j=1}^{\min(Q_i, K)} rel_{D_i}(R_i(j))$
        \item $MAP =  \frac{1}{M} \sum_{i=1}^M \frac{1}{N_i} \sum_{j=1}^{Q_i}    \frac{rel_{D_i}(R_i(j))}{j}$
        \item $NDCG@K = \frac{1}{M} \sum_{i=1}^M \frac{\sum_{j=1}^n \frac{rel_{D_i}(R_i(j))}{\log(j+1)}}{IDCG(D_i, K)} $
    \end{itemize}

P@K measures the fraction of first K recommended items that the user did actually consume averaged on users, MAP measures the fraction of recommended items on the set of true relevant items, and NDCG@K measures similar to MAP but takes the rank of the recommendation into account. 

Here we choose K = 30. The results are shown in Table \ref{tab:res}. We find noticeable increases over the \textbf{SPARSE} baseline for all other four datasets enhanced with fine-tuning and/or prompt learning. We compute the Average Percent Improvements (Avg. Imp\%) for all three metrics over the baseline. Overall, \textbf{SENT-ROBERTA} achieves the highest performance in MAP and NDCG@30 (0.5999 and 0.6403 respectively), and has the highest average \% increase (30.74\%). \textbf{SENT-BERT} has the highest P@30 score (0.0518), while having the lowest MAP and NDCG@30 (0.4854 and 0.5409 respectively) among the enhanced RecSys. After incorporating Prompt Learning, \textbf{SENT-PROMPT-BERT} improves significantly over \textbf{SENT-BERT} in MAP and NDCG@30 scores (from 0.4854 to 0.5832 and 0.5409 to 0.6258 respectively) while its P@30 score decreases (from 0.0518 to 0.0392). However, \textbf{SENT-PROMPT-ROBERTA}'s performance drops when compared to \textbf{SENT-ROBERTA} on all three metrics. Its MAP score decreases from 0.5999 to 0.577, NDCG@30 from 0.6403 to 0.6196, and P@30 from 0.0399 to 0.0389. One potential explanation for this is that the Collaborative Filtering may not at its pure optimal configuration and thus is underfitting the specific \textbf{SENT-PROMPT-ROBERTA} dataset. However, from the results above, we do find that enhanced RecSys improves the baseline RecSys noticeably, indicating the enhancements are working pretty well on the RecSys.

\section{Conclusion}
In this project, we demonstrate that utilizing advanced pretrained models such as BERT and RoBERTa to predict sentiment rating based on text reviews greatly address the data sparsity issue in CF recommender. Prompt-based learning paradigm, although superior to traditional fine-tune paradigm in sentiment analysis, shows no advantage to further improve the CF recommender. We believe this conclusion needs to be tested on additional datasets in the future to avoid the serendipity of a single dataset. Besides, potential future work include optimizing the prompt design(e.g. the format of the template), and attempting on different prompt methods such as soft prompt.

\newpage
\clearpage

\section{GitHub Link} 
\url{https://github.com/hhhhzy/nlu\_project}

\bibliographystyle{acl_natbib}
\bibliography{acl_latex}
\nocite{*}

\newpage
\clearpage
\appendix

\section{Appendix}
\subsection{Prompt Learning Workflow}
\label{sec:appendix}

\end{document}